\documentclass[%
reprint,
superscriptaddress,
nofootinbib,
amsmath,amssymb,
aps,
]{revtex4-2}
\usepackage[T1]{fontenc}
\usepackage[utf8]{inputenc}
\usepackage{lmodern}
\usepackage{epsfig} 
\usepackage{graphicx}
\usepackage{amsmath}
\usepackage{amssymb}
\usepackage{ulem}
\usepackage{bbold}
\usepackage{slashed}
\usepackage{mathtools}
\usepackage{lipsum}
\usepackage[compat=1.0.0]{tikz-feynman}

\tikzfeynmanset{
	fermion2/.style={
		/tikz/postaction={
			/tikz/decorate=false,
		},
	},
}
\tikzfeynmanset{
	fermion3/.style={
		/tikz/decoration={
			markings,
			mark=at position 0.5 with {
				\node[
				dot,
				fill,
				draw=none,
				] { };
			},
		},
		/tikz/postaction={
			/tikz/decorate=true,
		},
	},
}

\DeclareMathOperator*{\SumInt}{%
	\mathchoice%
	{\ooalign{\raisebox{.15\height}{\scalebox{0.9}{$\textstyle\sum$}}\cr\hidewidth$\displaystyle\int$\hidewidth\cr}}
	{\ooalign{\raisebox{.14\height}{\scalebox{.7}{$\textstyle\sum$}}\cr\hidewidth$\textstyle\int$\hidewidth\cr}}
	{\ooalign{\raisebox{.2\height}{\scalebox{.6}{$\scriptstyle\sum$}}\cr$\scriptstyle\int$\cr}}
	{\ooalign{\raisebox{.2\height}{\scalebox{.6}{$\scriptstyle\sum$}}\cr$\scriptstyle\int$\cr}}
}

\newcommand{\avg}[1]{\left\langle #1 \right\rangle}

\definecolor{dgreen}{rgb}{0.1,0.5,0.1}
\definecolor{lblue}{rgb}{0.2,0.35,1}
\definecolor{webred}{rgb}{0.75,0,0}

\begin{document}\preprint{APS/123-QED}
	
	\title{New Tool to Detect Inhomogeneous Chiral Symmetry Breaking}
	
	\author{Theo F. Motta}
	\affiliation{Institut für Theoretische Physik, Justus-Liebig-Universität Gie\ss en, 35392 Gie\ss en, Germany.\\ \,}
	\affiliation{Technische Universität Darmstadt, Fachbereich Physik, Institut für Kernphysik, Theoriezentrum, Schlossgartenstr.\ 2 D-64289 Darmstadt, Germany.\\ \,}
    \affiliation{Instituto de Física Teórica, Universidade Estadual Paulista, 01140-070 São Paulo, SP, Brazil}

    \author{Julian Bernhardt}
	\affiliation{Institut für Theoretische Physik, Justus-Liebig-Universität Gie\ss en, 35392 Gie\ss en, Germany.\\ \,}
	
	\author{Michael Buballa}
	\affiliation{Technische Universität Darmstadt, Fachbereich Physik, Institut für Kernphysik, Theoriezentrum, Schlossgartenstr.\ 2 D-64289 Darmstadt, Germany.\\ \,}
	\affiliation{Helmholtz Forschungsakademie Hessen f\"{u}r FAIR (HFHF), \\
		GSI Helmholtzzentrum f\"{u}r Schwerionenforschung,\\
		Campus Darmstadt,
		64289 Darmstadt,
		Germany. \\ \,}
	
	\author{Christian S. Fischer}
	\affiliation{Institut für Theoretische Physik, Justus-Liebig-Universität Gie\ss en, 35392 Gie\ss en, Germany.\\ \,}
	\affiliation{Helmholtz Forschungsakademie Hessen f\"{u}r FAIR (HFHF), \\
		GSI Helmholtzzentrum f\"{u}r Schwerionenforschung,\\
		Campus Gie\ss{}en,
		35392 Gie\ss{}en,
		Germany.}
	
	\begin{abstract}
		In this letter, we discuss a novel method to search for inhomogeneous chiral symmetry breaking in theories with fermions. 
		The prime application we have in mind is QCD, but the method is also applicable for other theories, including solid state 
		applications. It is based on an extension of the chiral susceptibility to inhomogeneous phases and it works as a stability
		analysis. Our method allows us to determine when homogeneous solutions have the tendency to create spatially modulated
		condensates. As proof of principle, we apply this technique to a rainbow-ladder QCD model
        and find that its phase diagram contains an inhomogeneous region.
        
	\end{abstract}
	
	\maketitle
	
	\section{Introduction}
	Chiral Symmetry Breaking ($\chi$SB) is one of the most consequential phenomena associated with strong interactions. 
	With massless quarks, the QCD Lagrangian is manifestly chirally symmetric, however the ground state breaks this symmetry. For two massless quark flavours, at high temperatures the system undergoes a second-order phase transition to the symmetric phase, which 
	becomes a cross-over at finite quark masses \cite{Borsanyi:2010bp,Bazavov:2011nk}. What happens to chiral symmetry 
	at large chemical potentials, however, is less clear. One possibility is that the system undergoes a transition to 
	a phase where chiral symmetry is broken \textit{inhomogeneously}, that is, the order parameter is a periodic function 
	of space. These phases have been seen in models of strong interactions \cite{Nickel:2009wj,Nakano:2004cd,Broniowski:2011ef,Buballa:2014tba}. 
	If real, these phases might occur in a region of the phase diagram that is connected to the physics of neutron stars 
	and neutron-star-mergers as well as with low-energy heavy-ion collisions. In fact, it has already been argued
    \cite{Pisarski:2021qof,Rennecke:2023xhc,Fukushima:2023tpv,Nussinov:2024erh} that moat regimes (a broader but correlated phenomenon to inhomogeneous phases, see also \cite{Fu:2019hdw,Haensch:2023sig}) would leave experimental 
	signatures in scattering experiments.
	
	From a theoretical perspective, it is highly non-trivial to obtain solid information on the existence of inhomogeneous
	phases at large chemical potential. Lattice calculations of QCD are notoriously difficult if not prohibited in this 
	region because of the sign problem \cite{Nagata:2021ugx}. Therefore, some attention has been paid to simpler (and partly
	lower-dimensional) models, such as Gross-Neveu or Nambu-Jona-Lasinio (NJL)
	type models \cite{Buballa:2020nsi,Pannullo:2021edr,Pannullo:2022eqh,Winstel:2022jkk,Pannullo:2023one,Pannullo:2024hqj,Thies:2006ti,Ciccone_2022,Koenigstein:2021llr,Koenigstein:2023yzv,Pannullo:2023cat,Buballa:2018hux,Buballa:2020xaa,PannulloPhysRevD.110.076006,Thies:2024anc},
	which are accessible to both, continuum and lattice methods. 
	
	In general, two strategies have been employed so far. Using a specific ansatz for the shape of the inhomogeneous phase, 
	one can calculate the free energy in this configuration and compare with the homogeneous one. One of the most commonly 
	used ansatzes is the so-called Chiral Density Wave (CDW) \cite{Nakano:2004cd}\footnote{The CDW is closely related to the (Quarkyonic) Chiral Spiral in $1+1$ dimensions \cite{Schon:2000qy,Kojo_2010}.
    Other commonly used ansatz functions are the Real-Kink-Crystal \cite{Thies:2006ti,Nickel:2009wj}, 
    the Twisted Kink Crystal \cite{Basar:2008ki,Basar:2009fg},    
    and similar hybrid condensates \cite{Nishiyama_2015}.}. Here, the condensates oscillate with spatial variable $\vec x$ 
    \begin{equation}\label{cdw}
     \begin{aligned}
         \avg{\bar\psi(x)\psi(x)} \propto \cos(2\vec{q}\cdot\vec{x}) \\
     \avg{\bar\psi(x)i\gamma_5\tau_3\psi(x)} \propto \sin(2\vec{q}\cdot\vec{x})
     \end{aligned}
    \end{equation}
    where $\tau_3$ is the third Pauli matrix in flavour space and $2\vec q$ is the wave vector of the modulation. 
    Alternatively, one
	can perform what is called a stability analysis which probes the stability of the homogeneous phase with respect to 
	very small inhomogeneous perturbations. This technique has the benefit that it is agnostic with respect to the 
	shape of the inhomogeneous phase. On the other hand, since it is based on small perturbations, 
 {it could miss instabilities with respect to large fluctuations.}
Nevertheless, these two techniques typically agree on second-order phase boundaries.
			
	\begin{figure}
		\centering
		\includegraphics[width=1\linewidth]{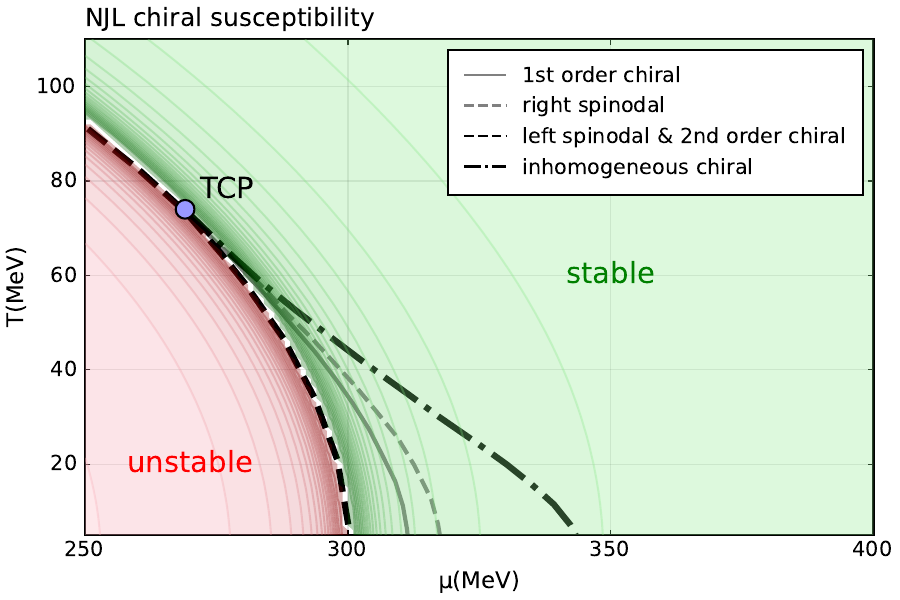}
		\caption{NJL example of the homogeneous chiral susceptibility $\chi_{\mbox{\tiny NJL}}$, 
		representing a probe for the chiral phase transition. 
        We show the phase diagram in the setup of Ref.~\cite{Nickel:2009wj}, see text 
        for details. Green shows positive values of $\chi_{\mbox{\tiny NJL}}$, 
        calculated with Eq.~(\ref{chinjl}). 
        Values in red show negativity of $\chi_{\mbox{\tiny NJL}}$.
			The dashed black line represents the second-order transition line at small chemical potential and the left 
			spinodal of the first-order transition region for $\mu > \mu_{\mbox{\tiny TCP}}$. The light gray lines show the first-order phase transition (solid) 
			and the right spinodal (dashed). The dashed-dotted black line shows the second-order transition to the inhomogeneous 
			phase found in Ref.~\cite{Nickel:2009wj}. The spinodals converge on the Tri-Critical Point (TCP).
            }\label{fig:njl}
	\end{figure}
	
    Although much has been learned from models, in the end 
	one needs to study the problem within QCD.
	Both, the ansatz and the stability analysis techniques face severe additional difficulties in QCD compared to their 
	application	in simple models. Due to the non-locality of the self-energy the expectation values of quark bilinears 
	are functions of two momentum variables in an inhomogeneous phase. To our knowledge, Ref.~\cite{Muller:2013tya} is
	the only attempt so far to deal with these difficulties in a functional approach to QCD via Dyson-Schwinger Equations (DSEs). 
	Using an ansatz for the quark propagator which yields condensates like Eq.~(\ref{cdw}), they show that spatially 
	inhomogeneous solutions to their set of (truncated) DSEs appear at high chemical potentials. 
	
	The stability analysis framework has only recently been generalised from model applications to QCD
	\cite{Motta:2023pks,Motta:2024agi}. Although agnostic in general, the QCD formulation of the framework relied on the specific shape of the perturbations (what is called the ``test-function'').
	Furthermore, so far, it has been possible to be applied to a very limited region of the phase diagram, namely the left 
	spinodal line. The details are explained in Refs.~\cite{Motta:2023pks,Motta:2024agi}, here we just note that the shape of 
{   the test-functions has constraints, which one can exactly enforce only in this region.
	Thus, although this version of the stability analysis is very promising and has led to 
	first evidence of a inhomogeneous phase in a simple truncation of QCD \cite{Motta:2024agi},
	it suffers somewhat from these limitations.
	
	
	In this letter, we overcome these shortcomings by introducing an alternative and novel technique 
	to perform a stability analysis. 
}
    The technique does not depend on any test-functions, and it is not restricted to any region of 
	the phase diagram. It is based on a well-known method to identify the second-order homogeneous phase boundaries, namely calculating 
	the chiral susceptibility (see e.g. Refs.~\cite{Qin_2011,Zhao:2006br}). However, by performing a space-dependent chiral rotation of the 
	quark fields, we are able to compute an \textit{inhomogeneous chiral susceptibility} and determine when the symmetric 
	solution is unstable with respect to inhomogeneous chiral symmetry breaking.
	

\begin{figure*}
	\centering
	\includegraphics[width=0.32\linewidth]{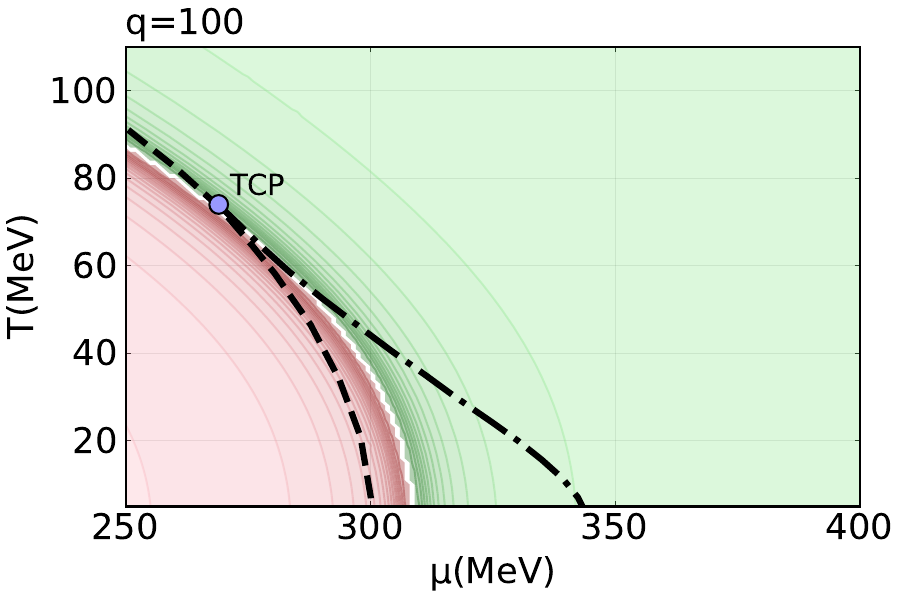}
	\includegraphics[width=0.32\linewidth]{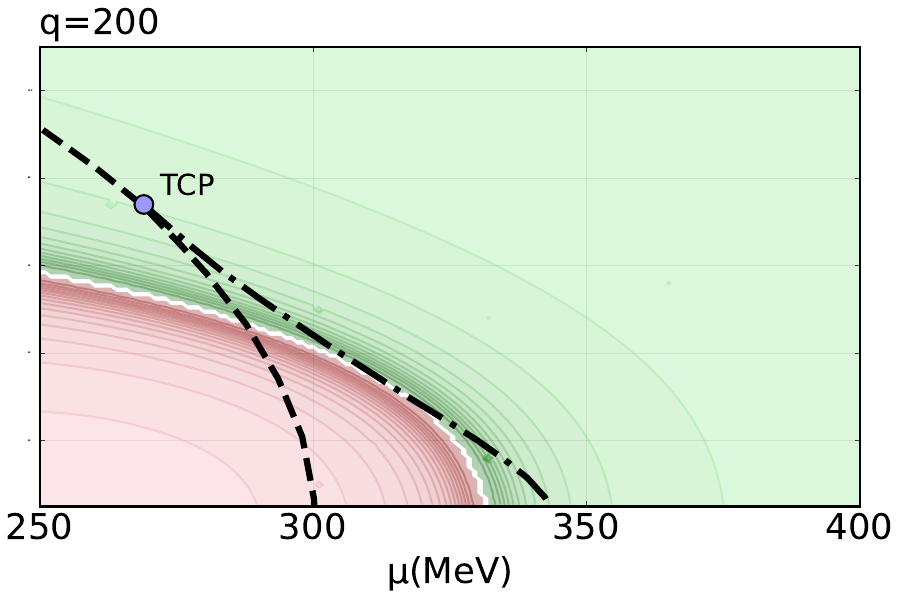}
	\includegraphics[width=0.32\linewidth]{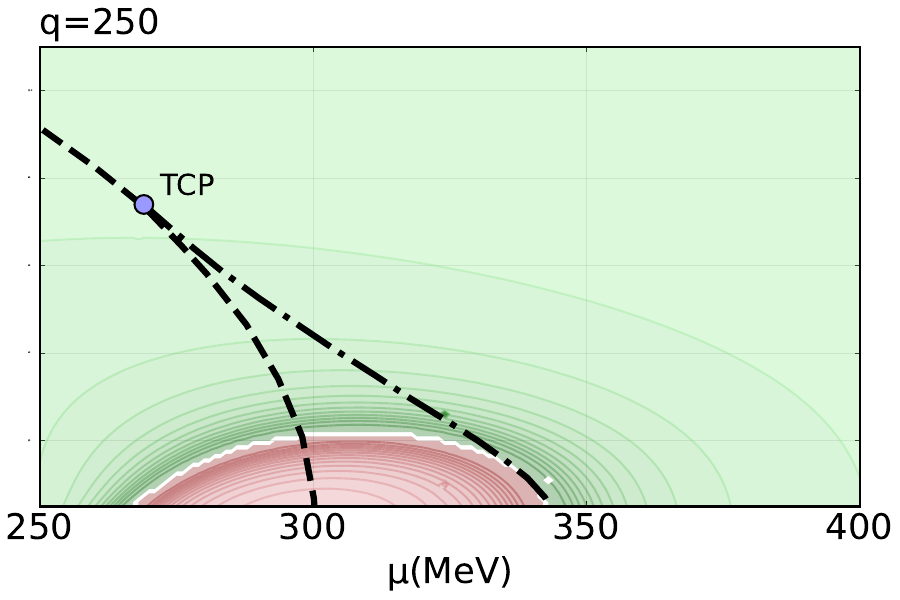}
	\caption{Inhomogeneous chiral susceptibility $\chi_{\mbox{\tiny NJL}}(q)$ for different values of $q$. In every diagram, the red region indicates instability of the symmetric phase with respect to the momentum scale $q$ indicated above the figures. Below the critical point, the boundary of the union of all regions where $\chi_\text{NJL}(q)$ is negative for some value of $q$ reproduces exactly the right boundary of the region where in Ref.~\cite{Nickel:2009wj} an inhomogeneous phase is found. The physically relevant result here is that we find negative $\chi$ for finite $q$ where for $q=0$ it is positive.}
	\label{fig:njl_q}
\end{figure*}

	\section{Inhomogeneous Chiral Susceptibility in NJL}
	Before we generalize it to QCD, let us illustrate the idea in the NJL model, defined by the Lagrangian\footnote{We will use the same Euclidian-space conventions as in Ref.~\cite{Motta:2023pks} and as for the numerical values for the couplings, cutoff, and regularisation scheme, we will take the setup of Ref.~\cite{Nickel:2009wj}.}
	\begin{equation}\label{njl_lag}
		\mathcal{L} =
		\bar{\psi}(\slashed\partial +m)\psi + G\Big(
		(\bar{\psi}\psi)^2 + (\bar{\psi}i\gamma_5\tau_a\psi)^2
		\Big).
	\end{equation}
    We will work in the chiral limit of zero bare mass $m$.
	In mean-field, the inverse dressed fermion propagator reads
	\begin{equation}
	S^{-1} = i\slashed p + M
	\end{equation}
	where $M$ is the dynamically generated mass, which depends on the chiral condensate and is an order parameter for $\chi$SB. One way to resolve the phase diagram of the theory is to calculate the derivative of this order parameter with respect to the bare quark mass.
 For the symmetric phase ($M\rightarrow 0$) and $N_c$ colors, this is simply
	\begin{equation}\label{chinjl}
	\chi_\text{NJL} = \frac{dM}{dm}= \frac{1}{1-16N_cG\SumInt_k 1/k^2},	
    \end{equation}
    where we denote the combination of the Matsubara sum of the frequencies $\omega_k$ and 3-momentum $\vec{k}$ integration as $\SumInt_k := T\sum_{\omega_k}\int d^3k/(2\pi)^3$. The susceptibility is negative where the symmetric phase is unstable with respect to 
    small homogeneous breakings of chiral symmetry, and it is positive otherwise.
    In a second-order chiral transition, $\chi_\text{NJL}$ changes its sign and diverges on the phase boundary. 
    Around a first-order transition, however, there is a region where both the restored and the broken solution are (meta-)stable. This is known as the spinodal region and its edges are known as the spinodal lines. 
    In Fig.~\ref{fig:njl}, we show how the sign of $\chi_\text{NJL}$ neatly resolves the second-order transition and left-spinodal lines in NJL, i.e., the regions where the restored solution 
    changes from unstable to (meta-)stable.
 
	Take now the following chiral-rotated quark fields \cite{Dautry:1979bk}
	\begin{equation}\label{rotation}
	\xi=e^{i\gamma_5\tau_3 q\cdot x}\psi.
	\end{equation}
	The NJL-Lagrangian, Eq.~(\ref{njl_lag}) with $m=0$, then transforms to 
	\begin{equation}\label{njl_lag2}
	\mathcal{L} =
	\bar{\xi}\Big(\slashed\partial + i\gamma_5\tau_3 \slashed q\Big) \xi + G\Big(
	(\bar{\xi}\xi)^2 + (\bar{\xi}i\gamma_5\tau_a\xi)^2
	\Big).
	\end{equation}
	For non-zero $q$, assuming a homogeneous condensate $\avg{\bar\xi\xi}={const.}$ and $\avg{\bar\xi\gamma_5\tau_3\xi}=0$, it is easy to show that one obtains a modulated $\avg{\bar\psi(x)\psi(x)}$ just like Eq.~(\ref{cdw}),
    \begin{equation}\label{psixi}
     \begin{aligned}
         \avg{\bar\psi(x)\psi(x)} = \cos(2{q}\cdot{x})\avg{\bar\xi\xi}, \\
     \avg{\bar\psi(x)i\gamma_5\tau_3\psi(x)} = \sin(2{q}\cdot{x})\avg{\bar\xi\xi}.
     \end{aligned}
    \end{equation}
	
	Under this transformation, the inverse fermion propagator becomes
	\begin{equation}
	S^{-1} = i\slashed p + M + i\gamma_5\tau_3 \slashed q.
	\end{equation}
	It can still be inverted and an inhomogeneous chiral susceptibility in the symmetric phase can be calculated\footnote{For this we temporarily re-introduce a small bare quark mass $m$ to the gap equation \textit{after} the chiral rotation and send it to zero after taking the derivative.} according to
	\begin{equation}\label{njlchi}
	\chi_\text{NJL}(q) = \frac{dM}{dm} = \frac{1}{1-16N_cG \SumInt_k \frac{k^2-q^2}{(k^2 +q^2)^2 -4(k\cdot q)^2}}.
	\end{equation}
    This expression is general, however, in practice we take the wave-vector $q_\nu$ taken to be fixed in the $z$ direction, i.e., $q_\nu=(0,0,q,0)$.
	In Fig.~\ref{fig:njl_q} we show how this resolves exactly the phase boundary of the inhomogeneous phase as well: 
	By varying $q$ we manage to find where the symmetric solution is unstable with respect to 
	inhomogeneous $\chi$SB, in regions where it is \textit{not} unstable with respect to homogeneous $\chi$SB. The union of the red regions in Fig.~\ref{fig:njl_q}, that is, regions of negative $\chi_\text{NJL}(q)$ for some value of $q$, reproduces exactly the unstable region in Ref.~\cite{Nickel:2009wj}. Note that the denominator of Eq.~(\ref{njlchi}) is exactly the stability condition found in standard stability analysis, so this correspondence is exact.
	
{   Having verified that our new method works and reproduces known results within the NJL-model, 
	let us now proceed extending this idea to QCD.
}	
	\section{QCD}

    {
    In QCD the setup is completely analogous. Under the chiral rotation, Eq.~(\ref{rotation}), the QCD Lagrangian
    \begin{equation}
        \label{qcd_norotation}
        \mathcal{L}_\text{QCD}=\bar{\psi} \left(\slashed D\right) \psi + \frac{1}{4}F_{\mu\nu}^aF_{\mu\nu}^a
    \end{equation}
    with covariant derivative $D_\mu$ and gluon field-strength tensor $F_{\mu\nu}^a$,
    transforms to
    \begin{equation}
        \label{qcd_rotated}
        \mathcal{L}_\text{QCD}=\bar{\xi} \left(\slashed D + i\gamma_5\tau_3\slashed q\right) \xi
        + \frac{1}{4}F_{\mu\nu}^aF_{\mu\nu}^a.
    \end{equation}

{As in NJL, the procedure is to first find the chiral symmetric solution for the quark propagator of the non-rotated system, described by Eq.~(\ref{qcd_norotation}). To this end we will use the DSE framework. Different approaches to obtain fully dressed n-point functions, as e.g. the 
    functional renormalisation group are equally suitable.
    In this letter, in order to illustrate our new method we employ a simple rainbow-ladder 
    truncation of the DSE for the quark propagator, given by}
	\begin{equation}\label{qDSE}
	\centering%
	\begin{tikzpicture}
	\begin{feynman}
	\vertex (a);
	\vertex [right=of a] (c);
	\vertex [right=0.5cm of c] (d);
	\vertex [above=0.4cm of c] (m1);
	\diagram*{
		(a) -- [fermion3] (c)
	};
	\draw (d) node { \({=}\)};
	\draw (m1) node { \(^{-1}\)};
	\vertex [right=0.5cm of d] (a1);
	\vertex [right=of a1] (c1);
	\vertex [right=0.5cm of c1] (d1);
	\vertex [above=0.4cm of c1] (m12);
	\diagram*{
		(a1) -- [fermion2] (c1)
	};
	\draw (d1) node { \({+}\)};
	\draw (m12) node { \(^{-1}\)};
	\vertex [right=0.5cm of d1] (a2);
	\vertex [right=0.55cm of a2] (gl1);
	\vertex [right=1cm of a2] (b2);
	\vertex [above=0.4cm of b2] (gldot);
	\vertex [right=1cm of b2] (c2);
	\vertex [left=0.5cm of c2] (gl2);
	\diagram*{
		(a2) -- [fermion2] (b2) -- [fermion2] (c2);
		(gl1) -- [boson,half left] (gl2)
	};
	\draw (gldot) node [gray, dot];
	\draw (b2) node [dot];
	\draw (gl2) node [];
	\end{feynman}
	\end{tikzpicture}\,.
	\end{equation}
	Here, the bare line denotes the (inverse) bare quark propagator, whereas the line with the black dot represents the (inverse) dressed quark propagator including quantum fluctuations.
    It has the general structure     
	\begin{equation}
	    S^{-1}(p) = i\vec{\slashed p} A(p) +  i \gamma_4 \tilde \omega_p C(p) + B(p)
	\end{equation}
    with two chiral-symmetry preserving terms parametrized by dressing functions $A$ and $C$
    (where $\tilde{\omega}_p = \omega_p + i \mu$) and a symmetry-breaking dressing function $B$ 
{   which vanishes in the symmetric phase.
	The self-interaction diagram contains two bare quark-gluon vertices. 
	The dressing of one of these vertices has already been combined with the dressing of the 
	gluon propagator (wiggly line) to an effective running coupling. For this coupling we will 
	use the Qin-Chang model \cite{Qin:2011dd} in the same setup as detailed in \cite{Motta:2024agi}. 
    This type of model has been applied to aspects of the QCD phase diagram before, see e.g. the review Ref.~\cite{Fischer:2018sdj}.
}    
    

    We now induce the inhomogeneous modulation by the chiral rotation.
    In an ansatz-based approach, one would have to go through
	the complicated and extremely costly exercise of solving the resulting set of DSEs self-consistently to get the stable 
	CDW solution.}
{    On the contrary, it is much simpler and numerically cheaper to determine the 
    inhomogeneous chiral susceptibility $\chi(\omega_p, \vec p \,| q)$.
Proceeding analogously to the NJL case yields the following expression (in the symmetric phase)
}	\begin{widetext}
		\begin{equation}\label{QCD}
		\chi(\omega_p, \vec p \,| q) =
		\frac{\partial B}{\partial m}
		=
		 1 + \frac{4}{3}Z_2^2\SumInt_k \frac{\chi(\omega_k,\vec{k} \,|q)\left(A(\omega_k ,\vec{k})^2\vec{k}^2 + C(\omega_k ,\vec{k})^2\tilde{\omega}_k ^2-q^2\right)}{\left(A(\omega_k ,\vec{k})^2\vec{k}^2 + C(\omega_k ,\vec{k})^2\tilde{\omega}_k ^2 +q^2\right)^2 -4\left(A(\omega_k ,\vec{k}) \vec k\cdot \vec q\right)^2} \times \frac{
			4\pi\left( \alpha_L\big((k-p)^2\big)+
			2\alpha_T\big((k-p)^2\big)\right)
		}{(k-p)^2}.
		\end{equation}
	\end{widetext}
    where $\alpha_{T,L}$ are the transverse and longitudinal components of the effective running coupling of the Qin-Chang model (see Ref.~\cite{Motta:2024agi} for details).
    Again, a non-trivial \textit{modulated} 
   chiral condensate is indicated if $\chi$ goes negative at finite $q$.
    Comparing Eq.~(\ref{QCD}) with the corresponding one for the NJL-model, Eq.~(\ref{njlchi}), we find the slight complication 
    that the susceptibility now depends on the quark momentum $p_\nu=(\omega_p, \vec p)$
    and appears on both sides such that the equation has to be solved self-consistently. 
    Note, however, that the quark dressing functions $A(\omega,\vec{k})$ and $C(\omega,\vec{k})$ on the right hand side 
    do not change under the chiral rotation in the symmetric phase and are therefore the same as those 
    calculated without
    the extra $\gamma_5 \tau_a \slashed q$ term present in the theory. This is a subtle and important point. In this analysis, we do not need to solve the DSE with the extra term present. 
    Instead, we study whether the addition of $\gamma_5 \tau_a \slashed q$ induces the formation 
    of a Dirac scalar term to the quark propagator. 
If so, then the susceptibility diverges {at the second-order phase boundary} and an inhomogeneous phase will be formed.

    \begin{figure}
    	\centering
    	\includegraphics[width=1\linewidth]{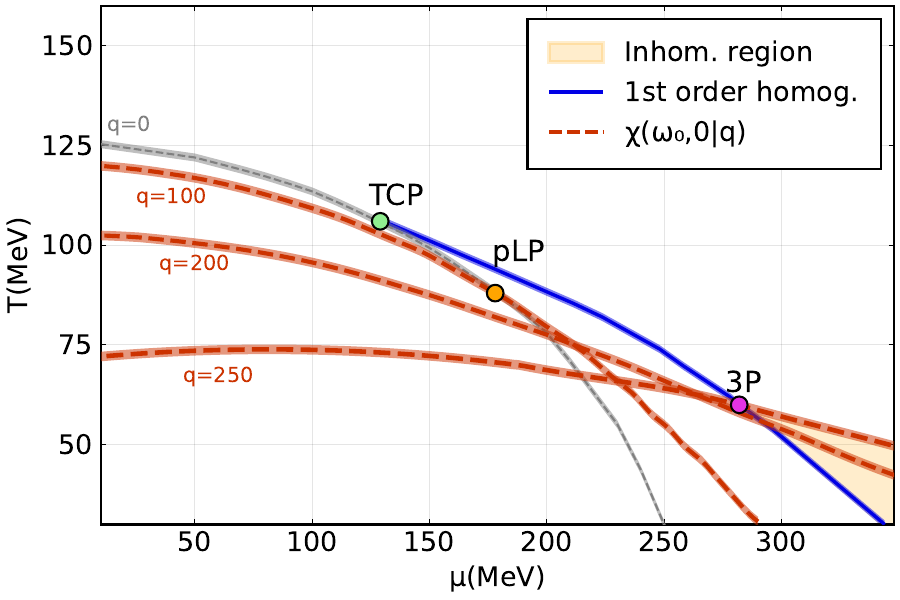}
    	\caption{Equivalent of Fig.~\ref{fig:njl_q} for our rainbow-ladder QCD model. Rather than showing a full heatmap of $\chi$, we show the lines in which $\chi$ diverges (dashed lines). The first-order chiral homogeneous phase boundary is shown by the solid line. The gray dashed line is the homogeneous chiral solution stability boundary, i.e., the combination of the second-order transition line with the left spinodal, which, as in Fig.~\ref{fig:njl}, agrees with the $q=0$ result of Eq.~(\ref{QCD}). 
        The highest point where the instabilities cross the first-order transition is our triple-point (3P) and the light shaded region shows where inhomogeneous phases appear. The proto-Lifschitz Point (pLP) agrees perfectly with Ref.~\cite{Motta:2024agi}
        All curves are shown with a width representing a systematic error (see Ref.~\cite{Motta:2023pks}).}
    	\label{fig:money}
    \end{figure}
    
	In Fig.~\ref{fig:money} we show the lines where the susceptibility diverges for different values of $q$. 
    %
    We confirm our finding from Ref.~\cite{Motta:2024agi} that near $T = 85$~MeV there is a ``proto-Lifschitz point'' (pLP), i.e., a point where the stability boundary of the symmetric phase w.r.t.\ inhomogeneous fluctuations meets the left spinodal of the homogeneous first-order transition. Below this point there is a region where the symmetric phase is unstable against developing inhomogeneous structures. 
    However, this instability is 
{without consequences} as long as it occurs to the left of the first-order homogeneous phase boundary (blue solid line), since in this region the symmetric phase is also disfavored against the chiral broken phase. 
    In Ref.~\cite{Motta:2024agi} the technical complexity of the applied test-function method prevented us from moving away very far from the left spinodal, and therefore we could only speculate about the existence of an inhomogeneous phase at higher chemical potential. 
    Our new method, on the other hand, can be applied at arbitrarily high chemical potential.     
    Varying the value of $q$ we find a triple-point (3P) at $T \simeq 60$~MeV where the instability line of the symmetric phase crosses the homogeneous first-order phase boundary. 
    In particular we establish that, while for temperatures above 60 MeV, the instability region does not extend beyond the homogeneous first-order transition, at lower temperatures it does. 
    Hence, in this regime the symmetric phase is unstable against inhomogeneous fluctuations in a region where it is favoured over the homogeneous chiral broken solution. This is an unambiguous proof that, in this truncation, a crystalline phase would surface.
    

Although this is very exciting, the main point of this study is to demonstrate that we now have a methodology where this analysis can be performed.
Our ultimate goal is to prioritize more realistic truncations as a direction for future work, where our methodology can provide more rigorous insights.

	\section{Conclusions}
	
	In this work, we proposed a new method to analyse the stability of a chirally symmetric phase against the formation
	of an inhomogeneous condensate via chiral symmetry breaking. The method is based on an extension of the concept of 
	chiral susceptibility {toward} probing the reaction of the theory to the presence of spatially modulated quark fields. 
	As a proof of principle, we showed that our method exactly reproduces previous NJL-model results \cite{Nickel:2009wj} on the presence of 
	inhomogeneous phases, we confirmed previous results for the presence of a proto-Lifschitz Point in a rainbow-ladder model 
	of QCD \cite{Motta:2024agi}, and we now also find a true instability of the symmetric phase in a regime where it is favored over the homogeneous broken phase. 

{ However, the main point of this letter is not this specific result but rather that the proposed method is indeed suited to perform such analyses. In fact, with this methodology we were able to obtain substantially more conclusive results than with the test-function based stability analysis applied in Ref.~\cite{Motta:2024agi}. 
{The latter requires numerically searching for a saddle point in the function's parameter space, 
which can be very expensive. Moreover, choosing the parameter space too small bears the risk 
of wrong results. 
For that reason this analysis is so far only feasible close to the left spinodal line \cite{Motta:2024agi} where the sufficiency of the parameter space can 
be ensured by analysing the homogeneous transition.}
%
Our new method requires no external parameters, and the \textit{only} real numerically costly operation is that of solving the homogeneous DSE.

The only restriction of the method is that, as it is based on a chiral rotation, it primarily signals instabilities toward CDW-like condensates, see Eq.~(\ref{psixi}). However, as known from model studies, even though the CDW is typically not the most favored inhomogeneous phase, in the chiral limit its second-order phase boundary to the symmetric phase coincides with those for other modulations. 
In this context  one should keep in mind that a stability analysis in general does not say much about the fully 
	self-consistent solution of the system. Here it just shows that there is the tendency to break chiral symmetry once we introduce a modulation, in this instance a CDW-like condensation. The actual 
	shape of the modulation would eventually to be determined by a self-consistent solution of the DSEs for all 
	dressing functions of the quark. This is a tremendous effort, but one that should be undertaken once our method 
	does show instabilities in more advanced truncations to QCD than used in this work. 
	

{
Some more care is required to generalise the method to be applicable away from the chiral limit. 
Then the chiral rotation (\ref{rotation}) is not a symmetry transformation of the QCD Lagrangian
but generates non-standard mass terms that need to be dealt with. This is work in progress. 
}

On the other hand, the new method can easily be generalized to any theory, including simple models, 
but also effective theories in solid state physics, to study the inhomogeneous breaking of any exact symmetry of that theory. For instance, in the QCD context it could also be used to study inhomogeneous color superconductivity \cite{Anglani:2013gfu}. Also, it is equally applicable to any truncation of QCD, with no computational costs in addition to solving the homogeneous DSE.
    Since it is not necessary to know the full momentum dependence of $\chi$,
	we expect this should also be 
    calculable with the functional renormalisation group. 
    

    Some studies suggest that additional quantum fluctuations could destabilise inhomogeneous condensates \cite{Pisarski:2020dnx,Lee_2015,fu2024qcdmoatregimerealtime,Kojo_2010}. In that case, we expect the homogeneous solution to the DSE to capture the energetically preferred, stable configuration. Other studies indicate that the boundary between inhomogeneous and restored phases may be first order \cite{Pisarski:2018bct}, limiting our ability to precisely determine the transition line. Nevertheless, our approach should still identify the corresponding spinodal.

	
	Finally, it is important to realise that such instabilities (or even their associated phenomenon, the moat regimes) are highly
	relevant as they might leave signatures in heavy-ion collisions \cite{Pisarski:2021qof}. Even if these phases turn 
	out not to be exactly stable in full QCD, these signatures might still be present due to long-lived meta-stable
	realisations of these phases as the system moves towards equilibrium \cite{Carlomagno:2018ogx}.

	\section*{Acknowledgements}
    It is our pleasure to thank Fabian Rennecke, Ricardo Costa de Almeida, Renan Hirayama, and Curtis Abell for helpful discussions about the content of this manuscript. This work has been supported by the Alexander von Humboldt Foundation, by the Funda\c{c}\~ao de Amparo \`a Pesquisa do Estado de S\~ao Paulo (FAPESP), grant 2024/13426-0, by the Deutsche Forschungsgemeinschaft
    (DFG) through the Collaborative Research Center TransRegio CRC-TR 211 ``Strong-interaction matter under extreme
    conditions'' and the individual grant FI 970/16-1. 

    \vfill
 
	\bibliographystyle{ieeetr}
	\bibliography{stabilitybib.bib}
	
\end{document}